\documentclass[a4paper,11pt]{article}
\usepackage{pos}
\usepackage{enumitem}

\title{Emergent phenomena from centre vortices in dynamical QCD}
\author*[a]{Waseem Kamleh}
\author[a]{James Biddle}
\author[a]{Derek B. Leinweber}
\author[a]{Adam Virgili}

\affiliation[a]{Centre for the Subatomic Structure of Matter, Department of Physics,\\ The University of Adelaide, SA 5005, Australia}

\emailAdd{waseem.kamleh@adelaide.edu.au}
\emailAdd{james.biddle@adelaide.edu.au}
\emailAdd{derek.leinweber@adelaide.edu.au}
\emailAdd{adam.virgili@adelaide.edu.au}

\abstract{Quark confinement and dynamical chiral symmetry breaking are two of the most important emergent properties of the theory of quantum chromodynamics. We review recent results studying centre vortices in $SU(3)$ lattice gauge theory with dynamical quarks. Through a vortex identification procedure, vortex-removed and vortex-only fields are obtained from the usual Monte Carlo generated gauge fields. Several comparisons between the untouched fields and the vortex-modified fields support the notion that centre vortices are fundamental to both confinement and dynamical chiral symmetry breaking in full QCD.}

\FullConference{%
  The 39th International Symposium on Lattice Field Theory (Lattice2022),\\
  8-13 August, 2022 \\
  Bonn, Germany 
}

\usepackage{slashed}
\newcommand{\muhat}{\hat{\mu}}
\newcommand{\nuhat}{\hat{\nu}}

\usepackage{tikz,varwidth}
\usepackage{tikzscale}
\usetikzlibrary{shapes.geometric, patterns}
\usetikzlibrary{arrows,arrows.meta,shapes,automata,petri,positioning,calc,decorations.markings}

\definecolor{purple}{HTML}{7570b3}
\definecolor{red}{HTML}{9E1B42}
\definecolor{orange}{HTML}{d95f02}
\definecolor{green}{HTML}{1b9e77}

\begin{document}
\maketitle

Emergent phenomena arise when a number of simple entities form complex collective behaviours as a response to their environment. There is extensive evidence that quark confinement and dynamical chiral symmetry breaking emerge from topological structures present in the nontrivial QCD vacuum known as centre vortices~\cite{tHooft:1977nqb,tHooft:1979rtg,DelDebbio:1996lih,Faber:1997rp,DelDebbio:1998luz,Bertle:1999tw,Faber:1999gu,Engelhardt:1999fd,Engelhardt:1999xw,Engelhardt:2000wc,Bertle:2000qv,Langfeld:2001cz,Greensite:2003bk,Bruckmann:2003yd,Engelhardt:2003wm,Boyko:2006ic,Ilgenfritz:2007ua,Bornyakov:2007fz,OCais:2008kqh,Engelhardt:2010ft,Bowman:2010zr,OMalley:2011aa,Trewartha:2015ida,Trewartha:2015nna,Greensite:2016pfc,Trewartha:2017ive,Biddle:2018dtc,Spengler:2018dxt,Biddle:2019gke}

\begin{figure}
\centering%
\begin{tikzpicture}[x=0.5cm,y=0.5cm,scale=1.0]
\draw [white, fill] (0,0) -- (10,0) -- (10,10) -- (0,10) -- (0,0);
\draw [black] (0,0) -- (10,0) -- (10,10) -- (0,10) -- (0,0);
\draw [thick, purple, ->, fill=purple!20] (5,5) circle (2.5);
\draw [dashed] (7,5) ellipse (1cm and 0.5 cm);
\draw [dashed, ->] (6.9,4) -- (7.1,4);
\node [blue] at (5,5) {$\bullet$};
\node [green] at (9,5) {$\bullet$};
\end{tikzpicture}%
\caption{Illustration of a Wilson loop (shaded circle) that is topologically linked to a centre vortex (dashed line). The representation of the Wilson loop lies within the plane of the page. The representation of the vortex has been sliced from a closed two-dimensional surface in four-dimensions into a loop in the perpendicular plane, piercing the page at the blue and green points.}
\label{fig:vortex}
\end{figure}
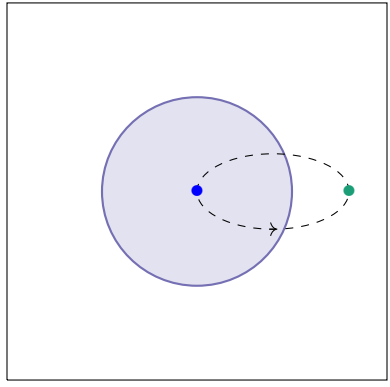%

In four dimensions, a thin centre vortex is a closed surface of centre flux within the gauge manifold. We say that a Wilson loop $W(C)$ along a curve $C = \partial A$ is topologically linked to the vortex surface if the vortex pierces the enclosed area $A$ only once. This concept is illustrated in Figure~\ref{fig:vortex}. The Wilson loop is in the plane of the page, with the enclosed area shaded in purple. As we cannot illustrate all four dimensions, the vortex surface is represented by the (oriented) dashed loop which lies in the plane perpendicular to the page. The blue dot shows the vortex piercing the area enclosed by the Wilson loop, and the green dot shows the same vortex piercing the same plane but exterior to the loop. As the vortex only pierces the Wilson loop once, the loop is topologically linked and $W(C)$ acquires a non-trivial centre phase $z,$ where
\begin{equation}
z = \exp\left(\frac{2\pi i}{3}m\right) I,\quad m \in \{ -1,0,1 \},
\end{equation}
is an element of the centre group of $SU(3).$

\begin{figure}[t]
  \centering
  \includegraphics[width=0.9\textwidth]{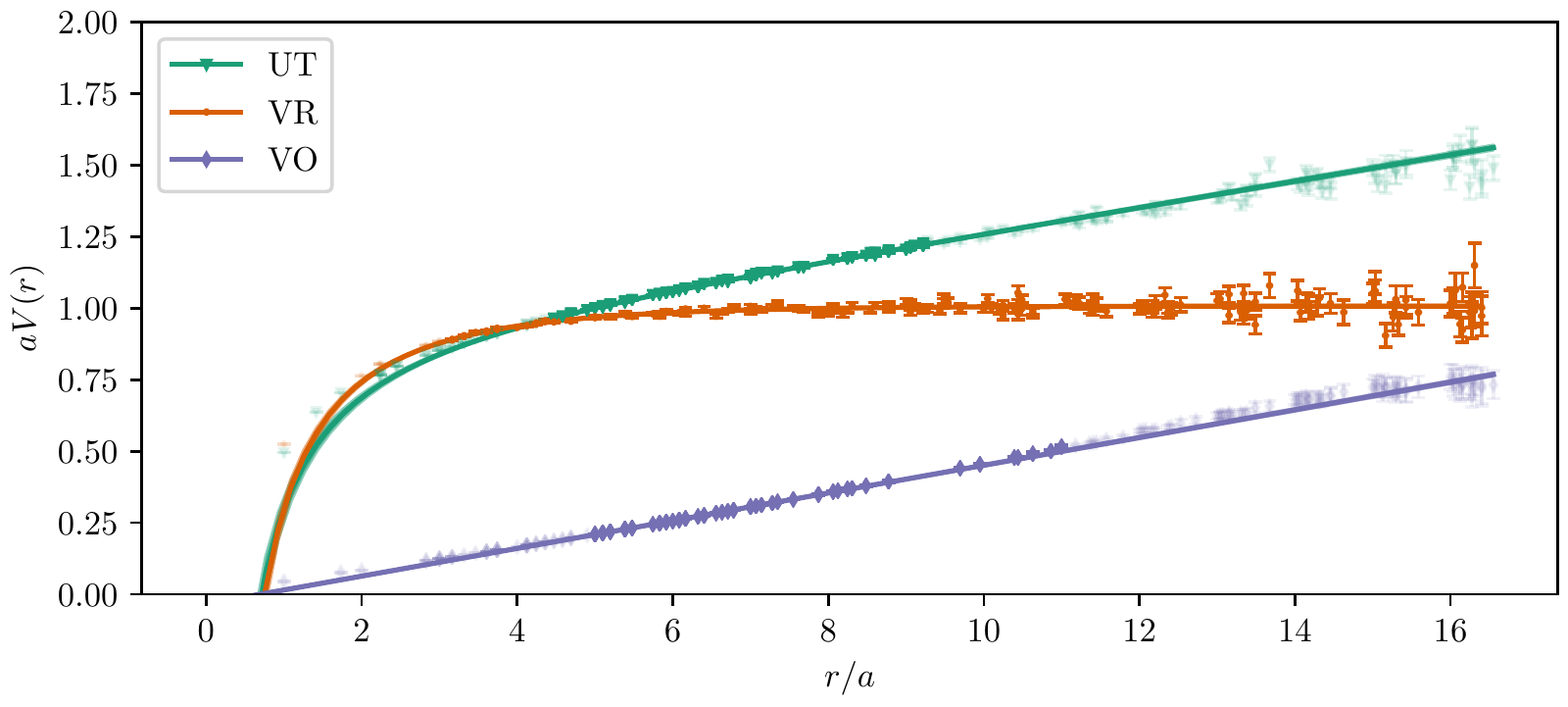}
  \caption{The static quark potential for the untouched (UT), vortex-removed (VR), vortex-only (VO) fields as calculated on $2+1$-flavour dynamical gauge fields with $m_\pi = 156\text{ MeV}.$ The solid points are included by the respective potential fits, using the fit functions described in the text.}
  \label{fig:sqplight}
\end{figure}

One of the advantages of lattice gauge theory is the ability to study vortex phenomology through the construction of \emph{vortex-modified} ensembles. This is done by decomposing the lattice gauge links $U_\mu(x)$ into the following form
\begin{equation}
U_{\mu}(x) = Z_{\mu}(x)\cdot R_{\mu}(x).
\end{equation}
The aim is to capture the vortex content in the field $Z_{\mu}(x),$ consisting only of centre elements. These are identified by first fixing the gauge field to Maximal Centre Gauge~\cite{DelDebbio:1996lih,Langfeld:1997jx,Langfeld:2003ev,OCais:2008kqh} and then projecting the links to the nearest centre element. Plaquettes with a nontrivial centre flux around the boundary are associated with vortices. These are the idealised \emph{thin} centre vortices describing the centre flux along a two-dimensional surface. This surface would have a finite profile in the case of the physical or \emph{thick} centre vortices. Nonetheless, there is a correlation between the identified thin vortices and the physical thick vortices, and we can gain insight about the latter by studying the former. The vortex-removed field $R_{\mu}(x)$ describes the remaining (short-range) fluctuations. Through the centre-projection process we create three distinct ensembles of SU(3) gauge fields:
\begin{itemize}
\item The original `untouched' configurations, $U_{\mu}(x),$

\item The projected vortex-only configurations, $Z_{\mu}(x),$

\item The vortex-removed configurations, $R_\mu(x) = Z^{\dagger}_{\mu}(x)\,U_{\mu}(x).$
\end{itemize}
Lattice studies comparing the three different ensembles and analysing the differences that emerge from the absence or presence of centre vortices grant insight into the role these structures play in QCD. There have been numerous vortex studies in pure gauge (i.e. Yang-Mills) theory. Herein we highlight published and upcoming results from the CSSM investigating vortices on $SU(3)$ gauge fields containing dynamical fermions, deferring to the relevant papers for the full details. 

It can be shown that when centre vortices percolate through the four-dimensional volume, this gives rise to an area law falloff for the Wilson loop, which then implies quark confinement~\cite{Greensite:2006ng} (at least for pure Yang-Mills theory). The precise meaning of quark confinement in the presence of dynamical fermions, which allows for string breaking, has been investigated in the context of $SU(2)$ gauge-Higgs theory (see for example~\cite{Fradkin:1978dv,Bertle:2003pj,Greensite:2003bk,Greensite:2006ng,Greensite:2017ajx,Greensite:2018mhh,Greensite:2020nhg}).
\begin{figure}[t]
  \centering
  \includegraphics[width=0.9\textwidth]{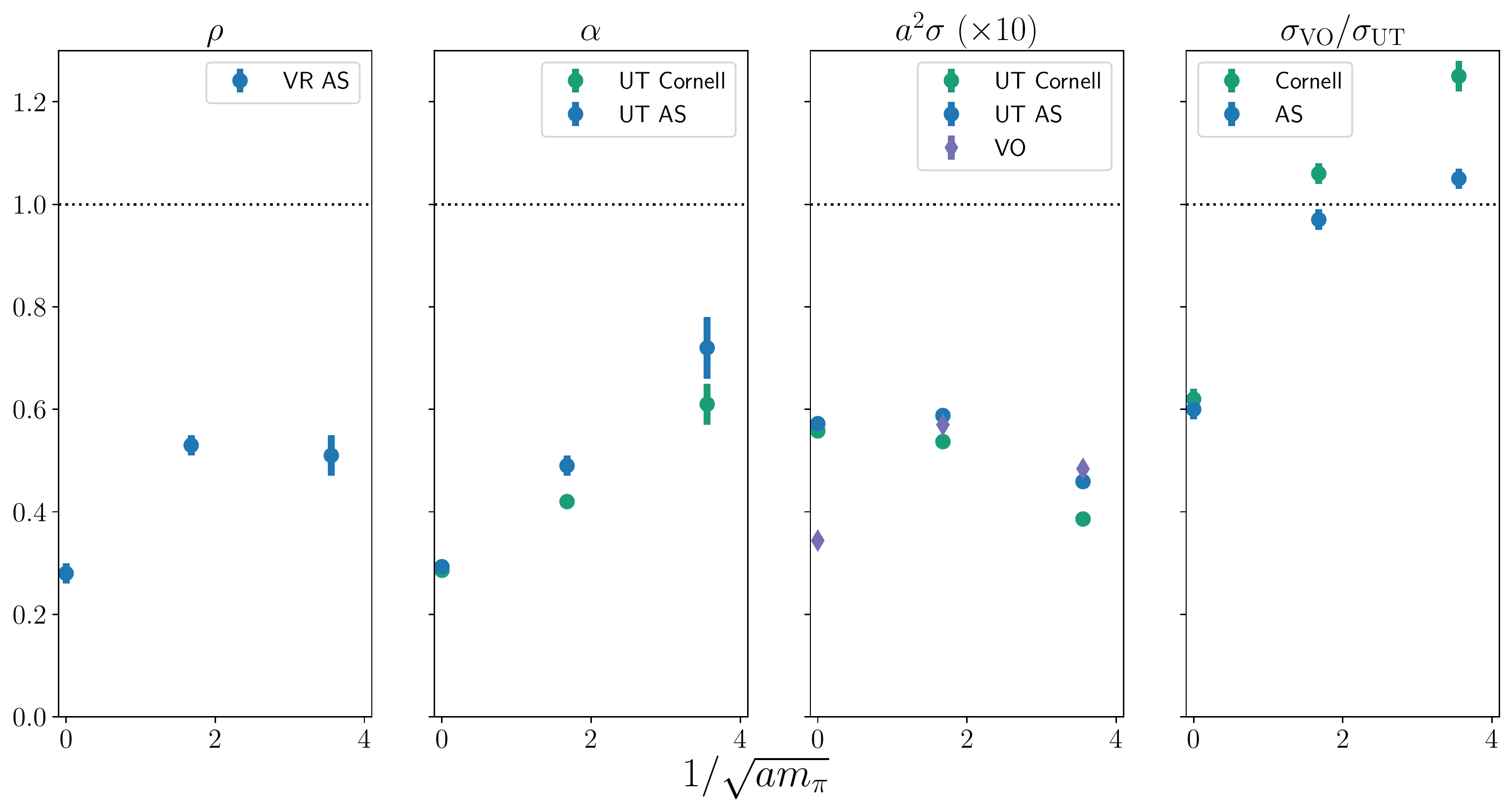}
  \caption{A summary of selected fit parameter values from Eqs.~(\ref{eq:volinear})-(\ref{eq:utcornell}) for the static quark potential results reported in~\cite{Biddle:2022zgw}. Results for the pure-gauge $(m_\pi = \infty)$, heavy dynamical $(m_\pi = 701\text{ MeV})$, and light dynamical $(m_\pi = 156\text{ MeV})$ gauge fields are plotted against $1/\sqrt{am_\pi}$ for presentation purposes. From left to right: the value of $\rho$ for the anti-screened vortex-removed potential fits; the value of $\alpha$ for the Cornell and anti-screened untouched potential fits; the value of the string tension $a^2\sigma$ for the Cornell and anti-screened untouched, and vortex-only linear potential fits; and the ratio of the vortex-only to untouched string tensions, with the latter obtained from the Cornell and anti-screened potential fits.}
  \label{fig:sqpsummary}
\end{figure}

As reported in~\cite{Biddle:2022zgw}, Figure~\ref{fig:sqplight} shows the static quark potential results for untouched, vortex-removed, and vortex-only fields on the light dynamical gauge field ensemble (i.e. the PACS-CS 2+1 flavour configurations~\cite{Aoki:2008sm} with $m_\pi = 156\text{ MeV}).$ The vortex-only field clearly generates a linear string tension and hence is fitted with a linear potential,
\begin{equation}
  V(r) = V_0 + \sigma\, r.
  \label{eq:volinear}
\end{equation}
By contrast, the vortex removed field is devoid of a confining linear potential~\cite{Langfeld:2003ev,OCais:2008kqh, Trewartha:2015ida} but retains a Coulombic interaction. In~\cite{Biddle:2022zgw} we find that a standard Coulomb term ansatz $V(r) = V_0 - \alpha/r$ does not describe the vortex-removed potential at moderate to large $r.$ This motivates using a modified Coulomb term to fit the vortex removed potential. Here we use an anti-screening potential,
\begin{equation}
  V_{\rm as}(r) = V_0 - \frac{\alpha}{ 1-e^{-\rho r}}.
  \label{eq:vras}
\end{equation}
At large $r,$ the effective coupling increases to create a constant potential $V_{\rm as}(r) \to V_0 - \alpha.$ At small $r, \tilde{\alpha} = \alpha/\rho$ becomes the effective Coulomb coefficient with $V_{\rm as}(r) \to V_0 - \tilde{\alpha}/r.$
The untouched potential is then fitted by fixing ${\rho}$ and adding the linear string tension,
\begin{equation}
  V(r) = V_0 - \frac{\alpha}{ 1-e^{-{\rho} r}} + \sigma\, r.
  \label{eq:utas}
\end{equation}
We note a Yukawa-style screening modification to the Coulomb term was also reported in~\cite{Biddle:2022zgw}, but we focus on just the anti-screened potential fits here for simplicity. % (both provide a similar quality of fit to the the vortex-removed data).

A summary of selected fit parameters plotted versus $1/\sqrt{am_\pi}$ for the pure gauge, heavy dynamical, and light dynamical ensembles is shown in Figure~\ref{fig:sqpsummary}. For the untouched results, we compare fit parameters from the anti-screened ansatz with those from the standard Cornell ansatz,
\begin{equation}
  V(r) = V_0 - \frac{\alpha}{r} + \sigma\, r.
  \label{eq:utcornell}
\end{equation}
Going from left to right, in the first plot we see that the value of $\rho$ on the pure gauge ensemble is much lower than for the two dynamical ensembles, which are roughly consistent. In the second plot, the value of the Columb term coefficient $\alpha$ for both the Cornell and anti-screened ans\"atze increases with decreasing quark mass, a possible indication of dynamical fermion screening. In the third plot we see that well-known issue that the vortex-only field recreates only $\sim 60\%$ of the untouched string tension in pure gauge theory. For the dynamical ensemble results, the vortex-only string tension agrees with the untouched Cornell potential fit at the heavy pion mass, but overshoots the untouched value at the light mass. However, when we instead use the anti-screened Coulomb term modification, the vortex-only and untouched string tensions on both dynamical ensembles are consistent. This view is reinforced by the fourth plot, which shows the ratio of the vortex-only to untouched string tensions for the Cornell and anti-screened ans\"atze. Similar results are reported in~\cite{Biddle:2022zgw} for the screened potential.

\begin{figure}[t]
  \centering%
  \includegraphics[width=0.48\textwidth]{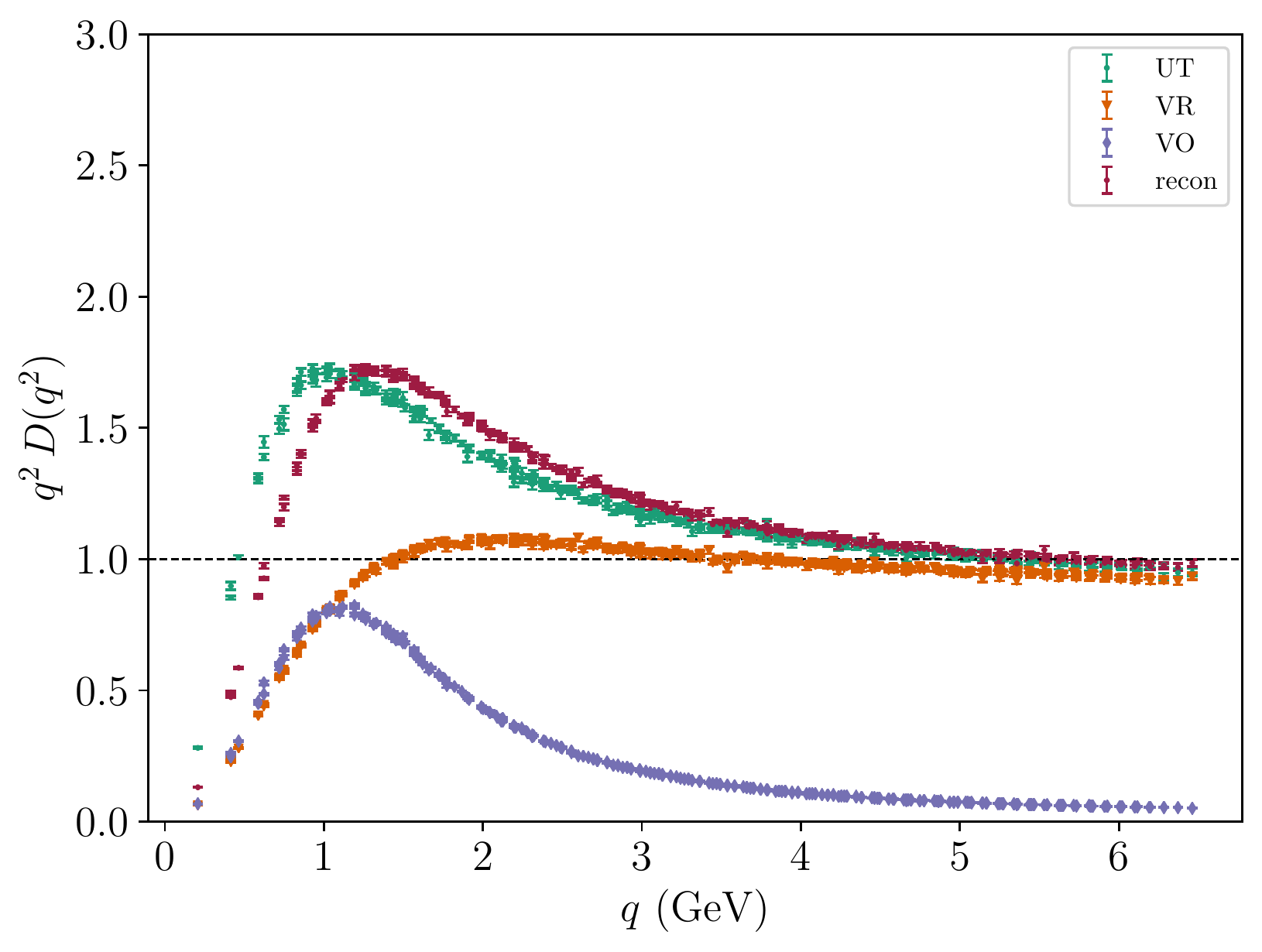}
  \includegraphics[width=0.48\textwidth]{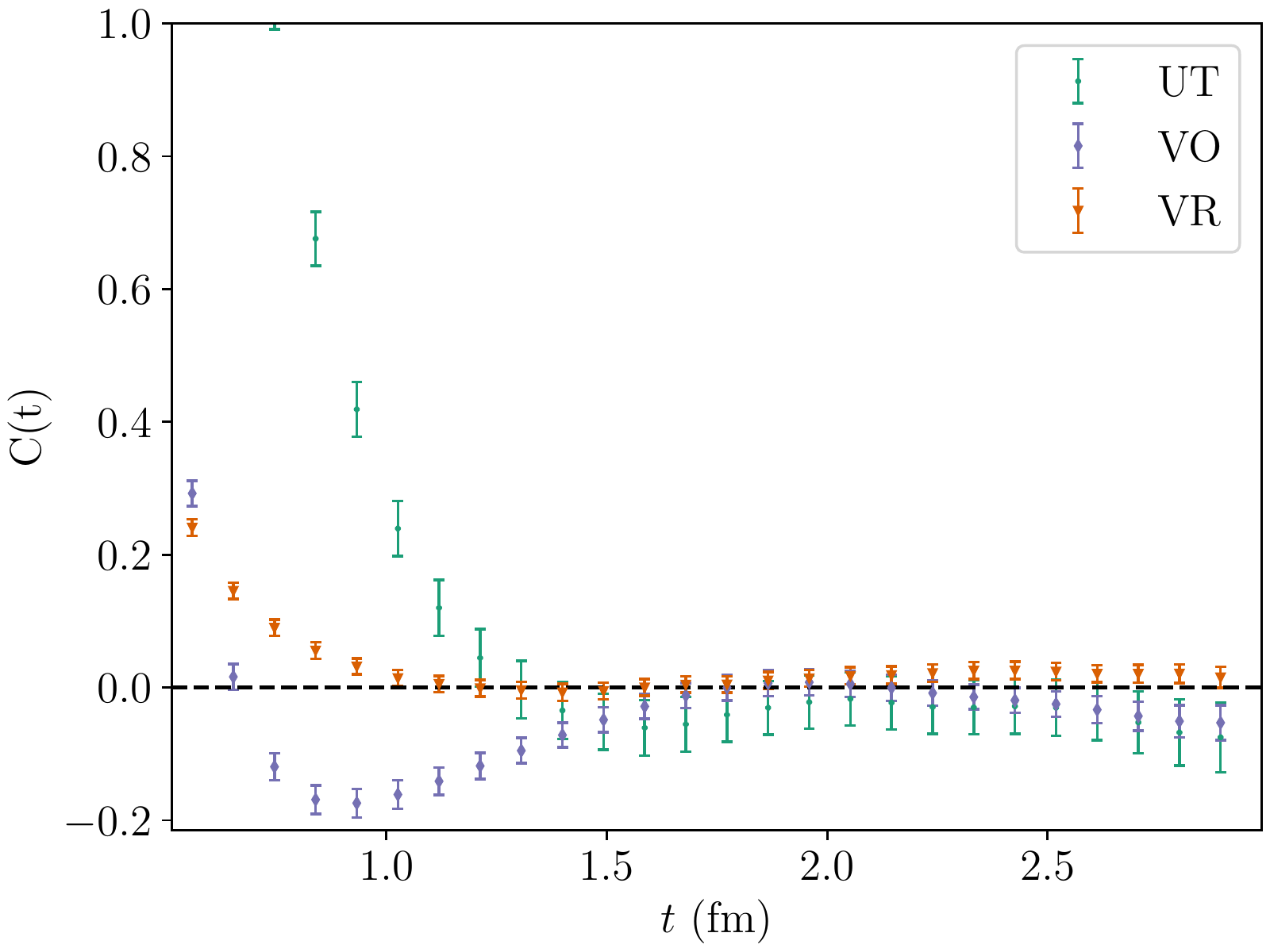}
  \caption{Gluon propagator results calculated in Landau gauge on the $2+1$-flavour dynamical gauge fields with $m_\pi = 156\text{ MeV}.$ The left plot shows $q^2 D(q^2)$ for the untouched (UT), vortex-removed (VR), and vortex-only (VO) fields. The reconstructed (recon) results are created from a linear combination of the vortex-only and vortex-removed propagators with coefficients obtained from a fit to the untouched propagator. The right plot shows the corresponding Euclidean correlator $C(t)$ described by Eq.~(\ref{eq:correlator}).}
  \label{fig:gluonprop}
\end{figure}
Figure~\ref{fig:gluonprop} shows two of the gluon propagator results reported in~\cite{Biddle:2022acd}. The nonperturbative scalar gluon propagator in momentum space is
\begin{equation}
  D(q^2) = \frac{Z(q^2)}{q^2},
\end{equation}
with $D(q^2) \to \frac{1}{q^2}$ at tree-level. The left hand plot shows the renormalisation function $Z(q^2) = q^2\, D(q^2)$ for the untouched and vortex-modified fields on the light dynamical ensemble. We see that vortex removal almost eliminates the infrared enhancement present in the untouched propagator. The vortex-only propagator shows significant infrared enhancement, capturing the long-distance physics. The reconstructed propagator formed from a linear combination of the vortex-only and vortex-removed propagators shows good agreement with the untouched results.

\begin{figure}[t]
\centering
\begin{tikzpicture}[x=1.0cm,y=1.0cm,scale=1.667]
\draw [black, fill=white] (-1.5,-1.5) -- (1.5,-1.5) -- (1.5,1.5) -- (-1.5,1.5) -- (-1.5,-1.5);
\draw [thin, gray, ->] (-1.5,0) -- (1.5,0);
\draw [thin, gray, ->] (0,-1.5) -- (0,1.5);
\draw [thin, black, dashed] (0,0) circle (1.0);
\node [black] at (1,0) {$\bullet$};
\node [blue] at (-0.5, 0.866) {$\bullet$};
\node [green] at (-0.5, -0.866) {$\bullet$};
\draw [thick, blue] (0,0) -- (cos 120, sin 120 );
\draw [thick, purple] (0,0) -- (-0.174, 0.985);
\draw [thick, purple] (0,0) -- (-0.766, 0.643);
\node [purple] at ( cos 100, sin 100) {$\bullet$};
\node [purple] at ( cos 140, sin 140) {$\bullet$};
\node [blue] at (-0.55, 1.02) {$\lambda$};
\node [purple] at ( 0.6, 0.5 ) {$\mp \omega(\lambda - \sigma)$};
\draw [thin, purple] (0.05,0.5) -- (-0.1,0.25); 
\draw [thin, purple] (0.05,0.5) -- (-0.2,0.22);
\draw[purple,dashed]([shift=(100:0.5cm)]0,0) arc (100:140:0.5cm);
\end{tikzpicture}
\caption{Illustration of the centrifuge preconditioning process described in the text.}
\label{fig:centrifuge}
\end{figure}
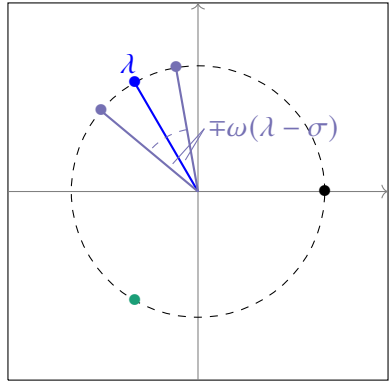
The right hand plot of Figure~\ref{fig:gluonprop} shows the Euclidean correlator,
\begin{equation}
  C(t)= \frac{1}{2\pi}\int_{-\infty}^{\infty}dp_0\, D(p_{0},\vec{0})\,e^{-ip_0\,t} = \int_0^{\infty}dm\, e^{-mt}\,\rho(m^2),
  \label{eq:correlator}
\end{equation}
calculated from the respective scalar propagators for the untouched, vortex-removed, and vortex-only fields. 
In order for the spectral representation to correspond to a physical particle, we require the spectral density $\rho(m^2)\ge 0.$ Hence, if $C(t) < 0$ for any $t$, then positivity is violated and gluons are confined.

We see immediately from the clear violations of spectral positivity that both the untouched and vortex-only fields are confining. By contrast, within statistical errors the vortex-removed correlator is non-negative and hence on the light dynamical ensemble we observe that vortex removal admits the possibility for gluons to be deconfined.

Given the importance of chiral symmetry in vortex studies, using the overlap fermion action is highly desirable due to its lattice-deformed chiral symmetry as represented by the Ginsparg-Wilson relation. The locality of the overlap fermion matrix is dependent on the underlying gauge fields satisfying a smoothness condition. The raw vortex-only fields are far too rough to meet this condition, so smoothing becomes essential when we wish to study the fermionic sector.

As reported in~\cite{Virgili:2022ybm}, applying standard smoothing techniques directly to the vortex-only fields fails. This can be understood intuitively by noting that there is no way to smoothly deform one centre element into another whilst remaining within the centre group. It is necessary to first perturb the vortex-only links away from the centre group before we can apply more traditional smoothing methods. The method we use to accomplish this perturbation is \emph{centrifuge preconditioning.}

Figure~\ref{fig:centrifuge} illustrates the centrifuge preconditioning process, which applies to the 3-vector of real phases in the non-compact representation of the centre element,
\begin{equation}
e^{i\lambda_\mu(x)} I \rightarrow [ \lambda_\mu(x), \lambda_\mu(x), \lambda_\mu(x) ].
\end{equation}
Initially the three phases are all equal. We break this symmetry by constructing the staple phase
\begin{equation}
  \sigma_\mu(x) = \frac{1}{6}\sum_{\nu \ne \mu} \left[ \lambda_\nu(x) + \lambda_\mu(x+\nuhat) - \lambda_\nu(x+\muhat) - \lambda_\nu(x-\nuhat) + \lambda_\mu(x-\nuhat) + \lambda_\nu(x-\nuhat+\muhat) \right]\,,
\end{equation}
and then selecting a pair of indices randomly for each link. This pair of phases is updated as follows
\begin{equation}
\lambda_\mu(x) \to (1\mp\omega)\,\lambda_\mu(x) \pm \omega\,\sigma_\mu(x),
\end{equation}
corresponding to a respective phase rotation by $\mp\omega(\lambda - \sigma).$ We set the centrifugal rotation angle to $\omega=0.02.$

\begin{figure}[t]
  \centering%
  \includegraphics[width=0.48\textwidth]{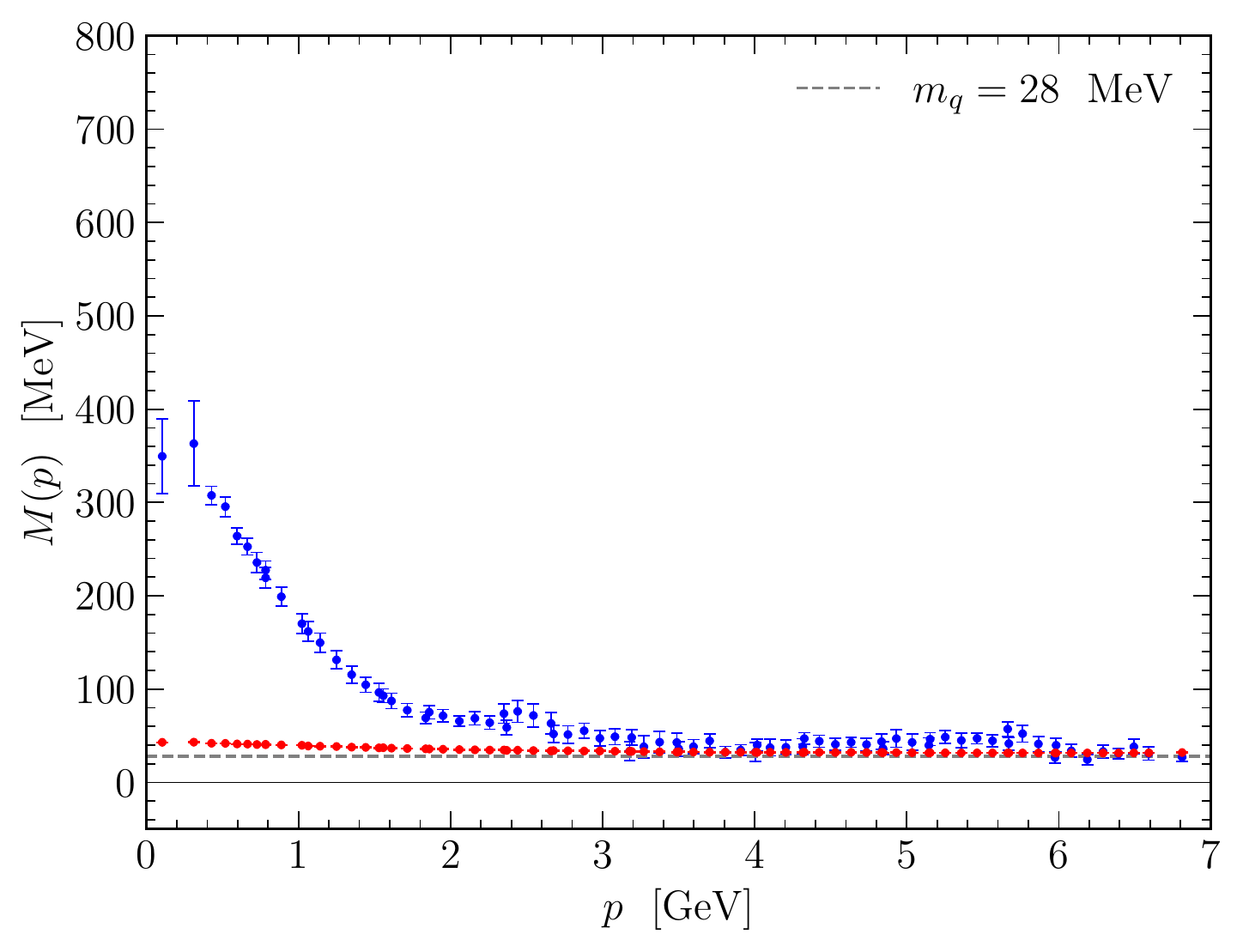}
  \includegraphics[width=0.48\textwidth]{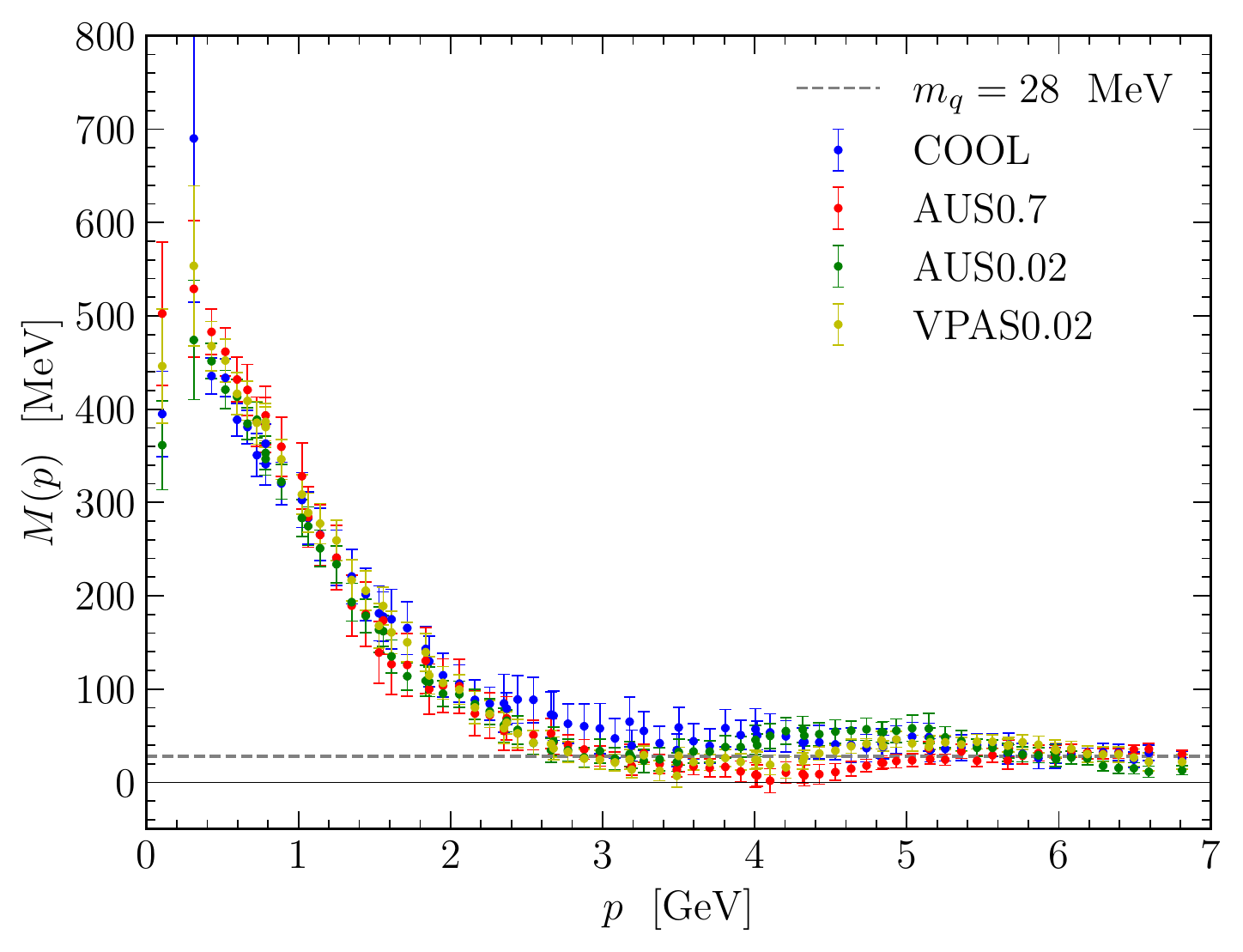}
  \caption{ The overlap quark mass function in Landau gauge at an input
    bare (valence) quark mass of $28$ MeV, as calculated on $2+1$-flavour dynamical gauge fields with $m_\pi = 156\text{ MeV}.$ The left plot shows the untouched (blue) and vortex-removed (red) results. The right plot shows the results on the vortex-only fields after smoothing has been applied as described in the text.}
  \label{fig:quarkprop}
\end{figure}
Figure~\ref{fig:quarkprop} shows the Landau gauge overlap quark propagator for the untouched and vortex-modified fields on the light dynamical mass ensemble. The quark propagator can be written as
\begin{equation}
  S(p) = \frac{Z(p)}{i\slashed{q} + M(p)},
\end{equation}
where $M(p)$ is the quark mass function, $Z(p)$ is the quark wave function, and $q$ is the kinematic tree-level momentum. At tree level, the mass function $M(p) \to m_q$ goes to the bare mass and the wave function $Z(p) \to 1$ goes to unity. The overlap quark propagator is able to be calculated directly on the untouched and vortex-removed propagators. Explicit chiral symmetry breaking is controlled by the bare quark mass $m_q.$ We see that vortex removal almost eliminates dynamical mass generation at the intermediate bare quark mass $m_q = 28\text{ MeV}$.

The right plot of Figure~\ref{fig:quarkprop} shows the vortex-only quark propagator after four different types of smoothing have been applied (see~\cite{Virgili:2022ybm,Virgili:inprep} for details):
\begin{itemize}[noitemsep]
\item Cooling
\item Annealed U smearing (AUS) with $\alpha=0.7$
\item Annealed U smearing (AUS) with $\alpha=0.02$
\item Vortex-preserving annealed smoothing (VPAS) with $\alpha=0.02.$
\end{itemize}
The three annealed smoothings are applied after centrifuge preconditioning. We see that the smoothed vortex-only field displays dynamical mass generation, and the asymptotic behaviour approaches the tree level value. These results will be reported in full in forthcoming work~\cite{Virgili:inprep}.

\section*{Acknowledgements}

We thank the PACS-CS Collaboration for making their 2+1 flavour configurations~\cite{Aoki:2008sm} available via the International Lattice Data Grid (ILDG)~\cite{Beckett:2009cb}.
This research was undertaken with the assistance of resources from the Pawsey Supercomputing Centre and the National Computational Infrastructure (NCI), provided through the
National Computational Merit Allocation Scheme and supported by the Australian Government through Grant No. LE190100021 via the
University of Adelaide Partner Share. This research is supported by Australian Research Council through Grants No. DP190102215 and DP210103706.
WK is supported by the Pawsey Supercomputing Centre through the Pawsey Centre for Extreme Scale Readiness (PaCER) program.

%\bibliographystyle{jhep}
%\bibliography{vortex_structure}

\providecommand{\href}[2]{#2}\begingroup\raggedright\endgroup

\end{document}